# A Dual Radiomic and Dosiomic Filtering Technique for Locoregional Radiation Pneumonitis Prediction in Breast Cancer Patients

Zhenyu Yang[1,2,*], Qian Chen[1,2,†], Rihui Zhang[1,2], Manju Liu[1,2], Fengqiu Guo[3], Minjie Yang[3], Min Tang[3], Lina Zhou[3], Chunhao Wang[4], Minbin Chen[3], Fang-Fang Yin[1,2]

[1] Medical Physics Graduate Program, Duke Kunshan University, Kunshan, Jiangsu, China
[2] Jiangsu Provincial University Key (Construction) Laboratory for Smart Diagnosis and Treatment of Lung Cancer, Kunshan, Jiangsu, China
[3] Department of Radiotherapy and Oncology, The First People's Hospital of Kunshan, Kunshan, Jiangsu, China
[4] Department of Radiation Oncology, Duke University, Durham, NC, United States

[†] Equally Contributed
[*] Corresponding authors:
Zhenyu Yang
Medical Physics Graduate Program
Duke Kunshan University, Kunshan, Jiangsu, China
E-mail: zy84@duke.edu

**Short Running Title: Explainable Dual-Omics Filtering for RP Prediction**

# Abstract

*Purpose:*

Radiation pneumonitis (RP) remains a critical complication of intensity-modulated radiation therapy (IMRT) for breast cancer patients, necessitating precise and explainable predictive models. This study developed a novel Explainable Dual-Omics Filtering (EDOF) model to integrate locoregional dosiomic and radiomic filtering for voxel-level RP prediction, with a focus on explaining the critical dosimetric and radiomic features contributing to RP development.

*Methods:*

This retrospective study analyzed data from 72 breast cancer patients treated with IMRT, of whom 28 developed RP. The EDOF model employs two key components: (1) dosiomic filtering to capture localized dose intensities and spatial distribution, and (2) radiomic filtering to extract texture-based features from pre-treatment CT scans. These features were subsequently analyzed using the Explainable Boosting Machine, a transparent machine learning algorithm that models non-linear relationships while maintaining feature-level explainability. Voxel-level RP predictions were validated through five-fold cross-validation using metrics such as area under the curve (AUC), sensitivity, and specificity. Feature importance was evaluated using mean absolute scores, and Partial Dependence Plots (PDP) provided detailed explanations of the response functions linking RP development to dual-omic features.

*Results:*

The EDOF model demonstrated high predictive performance, achieving an AUC of $0.95 \pm 0.01$ and sensitivity of $0.81 \pm 0.05$. Key features driving RP predictions included dosiomic *Intensity Mean*, radiomic *GLRLM-based SRLGLE*, and dosiomic *Intensity Mean Absolute Deviation*. PDPs revealed clinically relevant dose thresholds and complex relationships between tissue heterogeneity and RP risk. RP risk increases beyond a 5 Gy dose threshold and a rapid rise in the 10–30 Gy range, consistent with clinical guidelines of monitoring V5Gy, V10Gy, and V20Gy, Furthermore, structural heterogeneity captured by SRLGLE was shown to significantly influence RP in specific lung regions.

*Conclusion:*

The EDOF model advances RP risk prediction by integrating spatially specific dosimetric and radiomic features with explainable machine learning. It enables clinicians to identify high-risk lung regions and understand the rationale behind predictions. This capability supports personalized radiation therapy, potentially improving patient safety and treatment outcomes.

# 1. Introduction

Breast cancer remains a significant global public health issue, representing the most frequently diagnosed malignancy in women[1,2]. In 2022, approximately 2.3 million women were newly diagnosed, with an estimated 670,000 deaths attributed to the disease[2]. Current breast cancer management incorporates multimodal approaches, including surgery, chemotherapy, endocrine therapy, immunotherapy, and radiation therapy[1]. Advanced radiation treatment techniques, such as intensity-modulated radiation therapy (IMRT), allows precise tumor targeting while minimizing toxicity to surrounding tissues[3,4]. The clinical effectiveness of IMRT in the treatment of breast cancer is widely reported. Despite these advancements, radiation pneumonitis (RP) remains a serious complication of radiation therapy[5,6]. RP is characterized by alveolar septal edema, endothelial cell swelling, and vascular wall thickening within the radiation field, with CT imaging often showing hazy opacities and vascular thickening[7]. Clinically, RP presents as symptoms such as shortness of breath, cough, and fever and may progress to chronic stages, leading to fibrosis and impaired respiratory function[7,8]. Severe cases of RP are associated with significant quality-of-life impairments and the risk of irreversible lung damage. Furthermore, RP complicates radiation treatment of breast cancer, often necessitating therapy interruptions or termination. At present, it is not very clear how the dose distribution and intensity are associated with RP occurrence. Therefore, accurately predicting RP prior to radiation treatment is crucial for optimizing treatment planning and minimizing patient risk[5].

Traditional approaches to predicting RP have largely depended on dosimetric data derived from dose-volume histogram (DVH) of radiation treatment planning[9–11]. Dosimetric parameters, namely dose–volume indices (DVIs), including lung volume of receiving 10 Gy, 20 Gy, 30 Gy or more (V10Gy, V20Gy, V30Gy) and mean lung dose (MLD) are widely utilized in clinical and research settings to estimate radiation exposure risks[12]. While these metrics summarize dose distribution effectively, they lack spatial specificity. The extraction of statistical information from the 3D dose distribution, known as dosiomic features, has been explored[13,14]. Dosiomics features were calculated from the three-dimensional dose distribution to describe the detailed measurements of radiation intensity and its spatial distribution across lung tissue. Studies have demonstrated positive correlations between dosiomic features and the RP risks[13,15–19]. However, achieving high predictive accuracy using dosiomic features alone remains challenging due to inter-patient variations in anatomy and structural radiation response. Radiomic analysis has recently emerged as a complementary approach in RP prediction. Radiomic extracts high-dimensional quantitative features from medical images to capture tissue characteristics and heterogeneity[20–23]. Evidence suggests a correlation between lung CT texture features and RP occurrence[19,24]. Machine learning (ML) model that integrates radiomic features with dosimetric data (i.e., DVI or dosiomic features) thus has shown potential to enhance RP prediction accuracy, offering a comprehensive framework for identifying patients at increased risk[16,17,19,25,26].

Despite these advances, there remain significant limitations that hinder their clinical applicability and overall effectiveness. A primary limitation in existing research is the lack of emphasis on identifying specific high-risk regions within the lung that are more prone to RP. Most studies aggregate radiomic features from the entire lung or treatment area, along with dosiomic data from entire lung dose, to predict RP risk on a patient basis[19,25]. The lack of specific high-risk regions limits the clinical application in guiding more refined treatment planning. Another major challenge is that many machine-learning-based RP prediction models function as "black boxes". The key radiomic and dosiomic features driving the RP predictions often unknown or poorly understood[27–29]. This lack of transparency hampers clinical applicability, as the relationships between these features and RP development remain unclear, potentially limiting trust and implementation in routine practice. This lack of transparency hampers clinical applicability, as the relationships between these features and RP development remain unclear, potentially limiting trust and implementation in routine practice[27–29].

This study aimed to develop an Explainable Dual-Omics Filtering (EDOF) model to predict the locoregional risk of RP in breast cancer patients. The key innovations include (1) the introduction of novel dosiomic filtering (locoregional dosiomic feature extraction) and radiomic filtering (locoregional image radiomic texture analysis) techniques to assess RP risk at a finer spatial resolution, and (2) providing detailed explanations of the response functions linking RP development to dosiomic and radiomic features. To the best of our knowledge, this is the first study to achieve accurate spatial prediction of high-risk RP volumes while offering an explainable framework that integrates dosiomic and radiomic factors. The obtained explanation successfully proves that our model is driven by (1) features that are appropriate for clinical practice and (2) decisions that are clinically defensible. These advancements have the potential to improve clinicians' understanding of RP prediction mechanisms and enable more personalized radiation therapy strategies to reduce RP incidence and severity.

The main contributions of our work can be summarized as follows:
- The study proposes an innovative machine learning framework that integrates locoregional dosiomic and radiomic filtering to extract spatially resolved features from treatment dose distributions and pre-treatment CT images. This dual-filtering strategy enables the construction of a voxel-wise risk map, distinguishing it from existing patient-level RP prediction models.
- Unlike conventional black-box models, the proposed method employs EBM, a transparent ensemble model that maintains high predictive performance while preserving interpretability. EBM provides feature-level attribution through mean absolute scores and partial dependence plots (PDPs), enabling clinicians to understand the influence of individual dual-omic features on RP risk.
- Through feature selection and importance analysis, the model reveals that dosiomic Intensity Mean, radiomic GLRLM-based Short Run Low Gray Level Emphasis (SRLGLE), and dosiomic Intensity Mean Absolute Deviation are critical predictors of RP. The PDPs demonstrate dose-response relationships consistent with clinical guidelines (e.g., V5Gy, V20Gy) and uncover complex dependencies between lung heterogeneity and RP risk, offering clinically defensible, data-driven explanations at the voxel level.
- The EDOF model achieved the highest voxel-wise prediction accuracy among all evaluated models, with an average AUC of 0.95 and sensitivity of 0.81 across five-fold cross-validation. Comparative analyses against ablation variants and alternative models (including Gradient Boosting Trees, U-Net++, and DVI-based patient-level classifiers) confirm that the integration of both dual-omic features and explainability is essential for accurate and clinically reliable RP prediction.

## 2. Related Works

### A. Dose-based RP prediction

Extensive research has demonstrated that DVH-derived DVI—such as V5, V20, and MLD—are significantly associated with the risk of RP[30–34]. As a result, these parameters are widely utilized as dose constraints in clinical treatment planning and evaluation. Furthermore, various predictive models have been developed based on DVI to identify patients at elevated risk of RP[35–37].

Despite their widespread clinical acceptance, traditional DVI inherently provide a predominantly one-dimensional and aggregate representation of dose distribution, inadequately addressing spatial heterogeneity within lung tissues. Recent studies have thereby introduced the concept of dosiomics, which extracts high-dimensional features from 3D dose distributions, thereby offering enhanced characterization of both spatial and statistical dose patterns. For example, Liang et al.[38] demonstrated that a logistic regression model incorporating dosiomic features achieved superior predictive performance (AUC = 0.78) compared to models based solely on DVI (AUC = 0.68) or normal tissue complication probability (NTCP) scores (AUC = 0.74) using 70 patients treated with volumetric-modulated arc therapy (VMAT). Similarly, Adachi et al.[39] compared dosiomic-based, DVI-based, and hybrid models combining both approaches in a cohort of 247 lung cancer patients. Their findings revealed that the hybrid model achieved the higher predictive performance (AUC = 0.85), outperforming DVI only (AUC = 0.66) and dosiomic only (AUC = 0.84). Moreover, Palma et al.[40] reported notable differences in locoregional dose distributions between RP and non-RP patients, particularly in the lower lung and cardiac regions, based on an analysis of 178 patients treated with passive scattering proton therapy.

### B. Radiomic-based and multi-feature-combined RP prediction

Radiomics, defined as the high-throughput extraction of quantitative handcrafted features from medical images, enables non-invasive characterization of spatial heterogeneity in lung parenchyma both prior to and during RT (if repeated imaging is available)[20–23]. The radiomic approach has shown considerable promise in predicting RP. For instance, Du et al.[41] extracted radiomic features from cone-beam computed tomography (CBCT) images acquired at different stages of RT, reporting a moderate predictive capability with an AUC of 0.70. In another study,

Raptis et al.[42] evaluated multiple ML models—including logistic regression, support vector machines, random forests, and deep neural networks—to correlate lung radiomic features with patient-level RP incidence, achieving notable predictive performances (AUC range: 0.81-0.87).

While radiomics alone demonstrates potential predictive value, many recent studies have explored multi-modal feature integration, particularly the combination of radiomic and DVI and/or dosiomic features[16,17,19,25,26]. Multimodal models have demonstrated superior predictive capabilities by leveraging complementary information from diverse data sources and capturing complex feature interactions. For example, Su et al.[43] employed artificial neural networks combining clinical factors (e.g., age, sex, comorbidities) with DVI (e.g., MLD, V20, V5) in a cohort of 142 patients, achieving an AUC of 0.85. Similarly, Kapoor et al.[44], Jiang et al.[45], and Bourbonne et al[46]. reported substantial improvements in predictive performance by incorporating blood biomarkers, radiomic features, and DVI parameters into ML models such as support vector machines and convolutional neural networks, resulting in impressive predictive accuracies with AUCs of 0.91, 0.94, and 0.90, respectively. Extending these approaches further, Niu et al.[47] developed a multimodal prediction framework integrating radiomic features, DVI, dosiomic features, and cytokine data. Evaluations across various ML algorithms, including support vector classification, logistic regression, Gaussian and Bernoulli naive Bayes, k-nearest neighbors, decision trees, and random forests, demonstrated the highest predictive performance with an AUC of 0.94.

Collectively, these findings underscore the significant potential of multimodal ML models for accurately stratifying patient-specific RP risk, thereby contributing substantively to personalized radiotherapy planning. Nevertheless, few studies have attempted the precise localization of spatial RP distribution, and the integration of diverse multimodal features often amplifies the complexity and "black-box" characteristics inherent to advanced ML algorithms. The lack of explainability hinders the identification of key predictive features and their mechanistic relationships with RP development, thereby limiting clinical translation[27–29].

# 3. Materials and Methods

## A. Patient Data

A retrospective analysis was conducted on 72 breast cancer patients treated with IMRT between 2017 and 2021 from our hospital under an IRB protocol. Inclusion criteria were as follows: (1) treatment delivered using a Varian iX Linear Accelerator, (2) adherence to standard breast cancer treatment protocols, with prescribed doses of 45–50.4 Gy in 25–28 fractions for whole breast (to breast PTV) cases or 60–62 Gy with 5–10 additional fractions for tumor bed boost cases, (3) lung constraints meeting clinical standards (ipsilateral lung: V30Gy <10% and V20Gy <15%), and (4) availability of at least one follow-up CT scans within three months post-treatment. CBCT is used for treatment verification Among collected patients, 28 developed RP, including 23 cases of grade I and 5 cases of grade II, as detailed in Table I.

For each patient, pre-treatment simulation CT scans, follow up CT scans, and IMRT planning dose distributions (all patients were treated using 2-5 beams IMRT) were collected. Dose calculations were performed using the Eclipse™ Treatment Planning System (Varian Medical System, Palo Alto, CA) with the AAA algorithm (1-mm dose grid size). For patients with RP, affected regions were first identified and contoured on the follow-up CT images by experienced radiologist, and the contoured regions were aligned with planning CT images by an At present, it is not very clear how the dose distribution and intensity are associated with RP occurrence an experienced medical physicist using deformable image registration (Elastix Medical Image Registration Toolkit[48]), as illustrated in Figure 1. Initially, lungs at the treatment breast side were segmented from both sets of CT scans and identified affected regions (potential RP regions) were labelled. The registration provided voxel-wise RP segmentation labels, co-registered with the spatial coordinates of the planning CT lung volumes, serving as the ground truth for RP regions.

*Table 1. The details of the included patient information.*

| Characters | Patients with RP | Patients without RP |
|---|---|---|
| **Age (yrs)** | 50 (26–79) | 49 (30–68) |
| **Gender** | | |
|   Male | 0 | 0 |
|   Female | 28 | 44 |
| **Chemotherapy** | | |
|   Yes | 20 | 34 |
|   No | 8 | 10 |
| **Tumor location** | | |
|   Left | 14 | 21 |
|   Right | 14 | 23 |
| **Irradiation location** | | |
|   Breast, axillary, infraclavicular lymph | 13 | 18 |
|   Breast only | 15 | 26 |
| **Dose fractionations** | | |
|   60-61 Gy/28 fractions | 18 | 28 |
|   40-56 Gy/25 fractions | 9 | 14 |
|   Others | 1 | 2 |
| **Lung volume dose** | | |
|   V5Gy (cc) | 47.34±12.22 | 43.26±14.19 |
|   V10Gy (cc) | 27.72±6.86 | 25.48±7.86 |
|   V20Gy (cc) | 13.34±2.73 | 12.43±2.78 |
|   V30Gy (cc) | 8.85±1.87 | 7.93±2.12 |
|   MLD (Gy) | 9.27±1.73 | 8.58±1.90 |

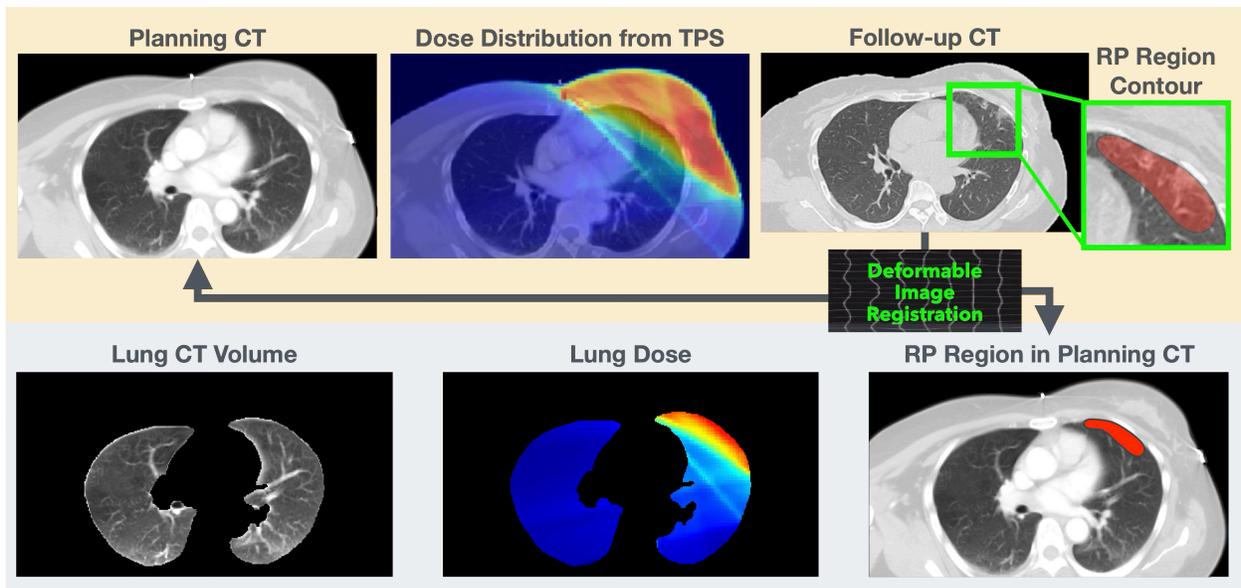

*Figure 1. An example of obtained patient data. For each patient, simulation CT scans prior to treatment, lung segmentation, and corresponding planning dose distributions were collected. The follow-up CT scans within 3 months post-treatment were obtained to identify the RP region.*

## B. EDOF Model Design

The overall design of the proposed EDOF model was summarized in Figure 2, which primarily included three components: (1) dosiomic filtering, (2) radiomic filtering, and (3) explainable voxel-wise RP prediction.

*B.1 Dosiomic Filtering*

As illustrated in Figure 2(A), the 3D dose distribution and corresponding lung segmentation for each patient were extracted from the treatment planning system. Dosiomic filtering applied a 3D sliding window kernel to systematically capture the regional dosiomic features across the entire regions of lung dose distribution[49,50]. Specifically, a pre-defined 3D kernel traversed the lung dose voxel by voxel, defining an isotropic sub-volume at each position. The dosiomic features were extracted from these cubic sub-volumes to characterize dose intensity and spatial distribution patterns, which transforming each voxel coordinate into a multi-dimensional feature vector. This process generated a set of 3D dosiomic filtering maps that could be visualized within the same reference frame as the original lung CT image. In this study, the kernel size for the 3D sliding window was set to $7 \times 7 \times 7$ mm³. Formally, for a lung volume with dimensions $I \times J \times K$ and $N$ dosiomic features, the filtering process produced $N$ dosiomic filtering maps, denoted as $\{\mathbf{D}^n\}_{n=1}^{N}$. Each map $D^n$ was represented as a 3D tensor $\mathbf{D}^n \in \mathbb{R}^{I \times J \times K}$, where the element $D_{i,j,k}^n$ corresponded to the value of the $n^{\text{th}}$ dosiomic feature at the $(i, j, k)^{\text{th}}$ coordinate.

Following previous dosiomic studies, a total of $n = 19$ dosiomic features were included. The full feature list can be found in Table S1 in *Supplementary Documents*. The multi-collinearity assessment was subsequently performed to remove the highly correlated dosiomic filtering maps. Specifically, Pearson correlation analysis was conducted for each pair of dosiomic maps, resulting in a covariance matrix for each patient. The hierarchical clustering was applied to the averaged correlation matrix (across the patients) to categorize the dosiomic features into well-separated clusters[51,52]. The representing feature was selected within each cluster as the independent feature, and a total of $P$ independent feature map $\{\mathbf{D}^p\}_{p=1}^{P}$ can be retained for the subsequent modeling.

*B.2 Radiomic Filtering*

The radiomic filtering technique adopted the similar conceptual design of the dosiomic filtering. As shown in Figure 2(B), the lung volume for each patient was first obtained based on planning CT and the lung segmentation. A 3D sliding window kernel was implemented to capture the local image texture intensity and texture characteristics within the lung CT. Similarly, given $M$ radiomic features, a total of $M$ radiomic filtering maps (denoted as $\{\mathbf{R}^m\}_{n=1}^{M}$) can be obtained. Each radiomic filtering map $\mathbf{R}^m$ is represented as a 3-dimensional tensor $\mathbf{R}^m \in \mathbb{R}^{I \times J \times K}$, where the element $R_{i,j,k}^m$ denotes the value of the $m^{\text{th}}$ radiomic feature at the $(i, j, k)^{\text{th}}$ tomographic coordinate.

Following previous lung radiomic studies and our pilot studies[49], a total of $m = 53$ radiomic texture features were included to comprehensively capture local texture characteristics within the lungs. The full feature list can be found in Table S2 in *Supplementary Documents*, which can be grouped into three types based on different joint-probability functions: (1) Gray Level Co-occurrence Matrix (GLCOM)-based, (2) Gray Level Run-Length Matrix (GLRLM)-based, and (3) Gray Level Size Zone Matrix (GLSZM)-based features. The multi-collinearity assessment was also performed, and $Q$ independent radiomic filtering maps $\{\mathbf{D}^q\}_{q=1}^{Q}$ can be obtained.

*B.3 Explainable RP Prediction*

As shown in Figure 2(C), the Explainable Boosting Machine (EBM) model was employed to predict voxel-wise RP occurrence[53,54]. EBM is an advanced form of Generalized Additive Model[53,54] that captures complex, non-linear relationships while maintaining explainability of individual features and the overall model decision-making. Mathematically, EBM is expressed as:

$$g(E_y) = \beta_0 + \sum f_i(x_i) \tag{1}$$

where $y$ is the outcome, $x_i$ is an input predictor feature, $f_i(x_i)$ represents the non-linear function for each predictor, and $\beta_0$ is the intercept. EBM learns these feature-specific functions using advanced techniques such as bagging and gradient boosting, with a constrained boosting procedure that trains one feature at a time in a round-robin manner with a low learning rate. As such, this approach reduces the influence of feature order and mitigates collinearity by iteratively refining each feature function independently.

Collectively, we proposed the EDOF model. For each tomographic coordinate $(i, j, k)$ within the lung, $p$ independent dosiomic features $D_{i,j,k}^p$ and $q$ independent radiomic features $R_{i,j,k}^q$ were combined into a $(p + q)$-dimensional feature vector. The combined vector was subsequently fed into an EBM, and a link function $g$ was employed to convert the EBM's continuous predictions into binary classification results. As such, the voxel-wise locoregional prediction of RP volume can be obtained.

EBM achieves explainability by representing predictions as a sum of feature-specific functions $f_i(x_i)$ rather than complex multi-layered structures. These function $f_i(x_i)$ enable the model to evaluate the impact of each feature on the outcome at both global and local levels. For global explainability, EBM calculates the mean absolute contribution of each function, known as the mean absolute score. These scores provide quantitative ranking of feature importance in predicting RP risk. For local explainability, EBM employs Partial Dependence Plots (PDPs) to visualize the relationship between individual features and model predictions. By plotting $f_i(x_i)$ against $x_i$, PDPs reveal subtle patterns, such as thresholds or non-linear dependencies, that influence risk assessment. For a more detailed technical description of EBM, refer to [53,55].

*B.4 Model Implementation and Evaluation*

A cohort of 72 breast cancer patients was divided into training and testing sets in an 8:2 ratio using a 5-fold cross-validation method. In each fold, 80% of patients were randomly allocated to the training set, while 20% constituted the testing set. This process was repeated across five iterations, ensuring that each patient appeared in both training and testing sets in different folds. Voxel-wise RP prediction results were assessed using sensitivity, specificity, accuracy, and area under the receiver operating characteristic curve (AUC). Performance metrics from all folds were aggregated to determine the model's overall effectiveness. Given that Grade-II RP patients typically exhibit larger RP volumes, the Dice Similarity Coefficient (DSC) was additionally calculated for this subset. Within each fold, the EBM's mean absolute scores identified the key radiomic and dosiomic features contributing to RP prediction, while PDPs illustrated the feature-specific relationships between these key features variations and the RP risk.

The radiomic and dosiomic filtering was implemented using an in-house developed toolbox with MATLAB (MATLAB R2023a; MathWorks, Natick, MA). For a more detailed technical description of filtering toolbox, refer to[49]. The EBM was implemented in a Python environment using the InterpretML library (Microsoft, Redmond, WA). The training process employed a logistic loss function, a boosting learning rate of 0.015, and a maximum of 25,000 boosting rounds, with early stopping triggered after 100 consecutive rounds without improvement to prevent overfitting. All computations were conducted on a workstation equipped with a 16-core Intel Core i7-13700KF CPU (3.4 GHz), 16 GB of RAM, and an Nvidia GeForce RTX 4070 GPU.

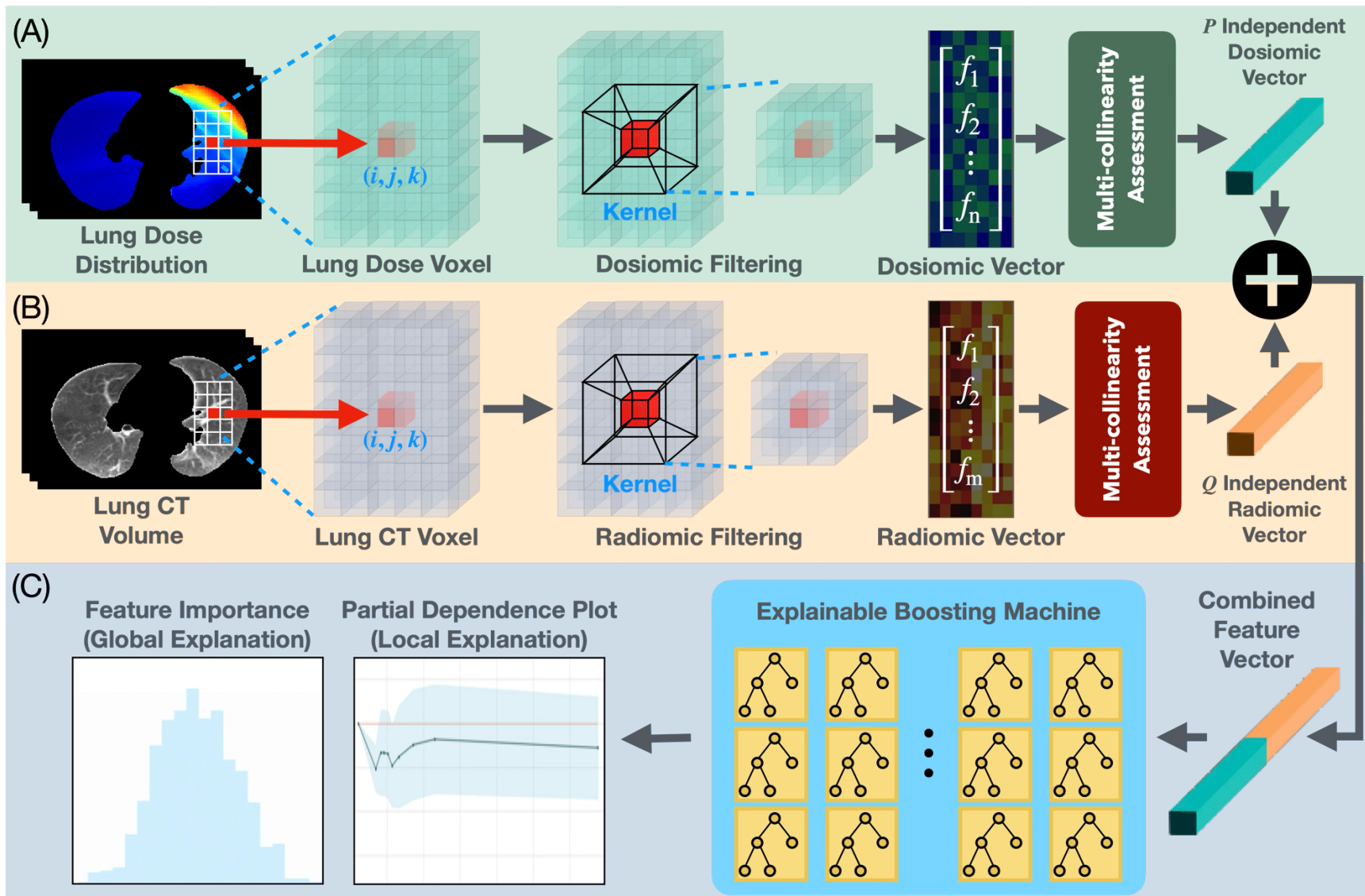

*Figure 2. The overall workflow of the proposed EDOF model, which compromise the three key components: (A) The dosiomic filtering component; (B) The radiomic filtering component; (C) The voxel-wise RP prediction and explanations based on the EBM model.*

## C. Ablation Study

Ablation studies were conducted to evaluate the contributions of each component of the EDOF model by systematically removing or modifying its components, resulting in three variants:

1) **EBM with Radiomic Filtering:** In this variant, the dosiomic filtering component was excluded and only $q$ independent radiomic features $R_{i,j,k}^q$ were used as input to an EBM model. This configuration assessed the impact of excluding detailed locoregional dosimetric information.
2) **EBM with Dosiomic Filtering:** In this variant, the radiomic filtering component was excluded and only $p$ independent dosiomic features $D_{i,j,k}^p$ were used as input to an EBM model. This variant evaluated the importance of radiomic features in capturing textural and structural variations relevant to RP risk.
3) **GBT with Dual-Omic Filtering:** In this variant, the EBM model was replaced with a Gradient Boosting Tree (GBT) model[56], while retaining both dosiomic $D_{i,j,k}^p$ and radiomic $R_{i,j,k}^q$ features. GBT is the state-of-the-art ML classifier[57]. This variant tested the prediction accuracy of the EBM and its ability to provide explainable predictions through feature-specific response functions.

Each variant was compared to the EDOF model to determine the relative contribution of each component. Training parameters, including the loss function, boosting learning rate, number of boosting rounds, and early stopping mechanism, were kept consistent across all variants. The same five-fold cross-validation scheme and training/testing set assignments were used to ensure comparability. Performance metrics, including sensitivity, specificity, accuracy, AUC, were statistically compared between each variant and the EDOF model using the Wilcoxon signed-rank test, with a significance threshold of 0.05.

**D. Comparison Study**

In the comparison study, two additional RP prediction models were also evaluated:
1) **U-Net++ Model:** The U-Net++ model[58] was implemented to predict RP on a voxel-wise basis. The input consisted of a two-channel combination of the original lung CT and the lung 3D dose distribution. U-Net++ employs multiple sub-networks of varying depths, connecting decoder stages through densely linked skip connections, which reduce semantic gaps between encoder and decoder layers. The encoder captures spatial features across multiple scales, while the decoder reconstructs these features into a fine-grained prediction map for voxel-wise RP prediction. This dense connectivity enhances sensitivity to subtle local variations compared to classic U-Net. Binary cross-entropy was used as the loss function, and the Adam optimizer was employed with an initial learning rate of $10^{-3}$ during model training.
2) **DVI Model** (patient-wise RP prediction): This model was designed to predict RP occurrence at the patient level rather than the voxel level. Following a traditional RP prediction framework, DVI—including GTV prescription dose, total lung dose, MLD, lung V5Gy, V10Gy, V20Gy, V30Gy, and mean heart dose—were extracted from the planning dose distribution and combined into a feature vector. This vector was input into a GBT model to estimate the probability of RP occurrence for each patient. During the training, trees number was set to 100, and binomial deviance loss was used with learning rate of 0.1.

The U-Net++ model was compared against the EDOF model in terms of voxel-wise performance metrics, using the Wilcoxon signed-rank test. For the DVI Model, patient-wise metrics, including sensitivity, specificity, accuracy, and AUC, were evaluated. Consistent five-fold cross-validation and training/testing set assignments were applied across all models.

## 4. Results

A total of 5 radiomic and 30 dosiomic were identified as the independent features after multi-collinearity assessment (the complete list can be found in Table S3 in *Supplementary documents*). Figure 3(A) provides a demonstration example of dual-omic filtered maps (including both radiomic and dosiomic) for a single patient. All feature maps were normalized to [0,1] range and overlaid on CT for better visualization. The dosiomic feature maps capture locoregional dose distribution patterns with distinct emphases. For example, dosiomic feature #1 highlights high-dose regions, while features #2 and #3 emphasize areas with significant dose variation or falloff. These maps provide a detailed spatial representation of dose distribution across lung tissue. In contrast, radiomic feature maps reflect textural variations within the lung. Radiomic features #1 and #2 identify heterogeneous regions associated with pulmonary vasculature, whereas feature #3 demonstrates higher values in lung with high air/tissue ratio. Together, these dual-omic filtering maps deliver a comprehensive spatial encoding of both dose and tissue characteristics.

Table I summarizes the performance metrics of the proposed EDOF model, its three ablation study variants, and two comparison models. Prediction performance was evaluated using five-fold cross-validation (mean±standard deviation), with "*" indicating statistically significant differences compared to the EDOF model. The EDOF model achieved the highest voxel-wise prediction performance, with a mean AUC=0.95±0.01, accuracy=0.93±0.02, sensitivity=0.81±0.05, and specificity=0.93±0.01. Given the relatively small RP volumes (compared to the entire lung), sensitivity is particularly critical for evaluating a model's ability to identify RP regions.

The EBM model with radiomic filtering alone demonstrated limited predictive performance, achieving an AUC=0.68±0.07 and sensitivity=0.01±0.01. In contrast, the EBM model with dosiomic filtering alone showed notable improvement, with AUC=0.94±0.01 and sensitivity=0.71±0.03. The GBT model with dual-omic filtering achieved AUC=0.95±0.01 and sensitivity=0.79±0.10, but remained inferior to our EDOF model in sensitivity. The U-Net++ model failed to capture voxel-wise RP volumes, demonstrating poor sensitivity=0.00±0.00 and specificity=1.00±0.00. Dice coefficients calculated for five Grade II patients further highlighted

these trends: the EDOF model achieved a DSC=0.78±0.07, while DSC=0.04±0.03/0.70±0.08/0.75±0.23/0.00±0.00 for the EBM with radiomic filtering only, dosiomic filtering only, GBT with dual-omics filtering, and U-Net++, respectively. Figure 3(B) illustrates an example of predicted RP volumes for a single patient across these models.

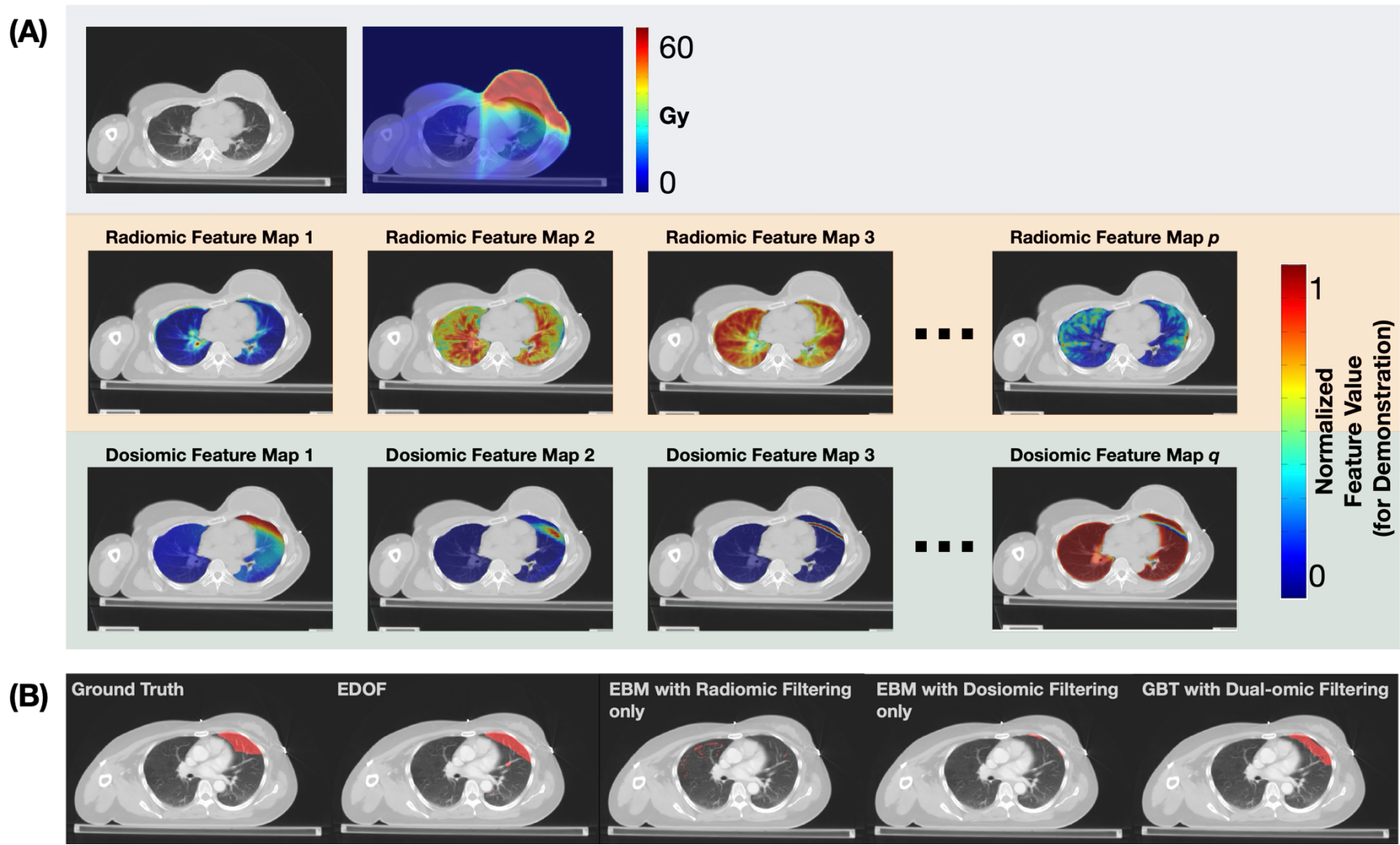

*Figure 3. (A) An example of the obtained dual-omic filtering feature maps for a single patient. The feature maps were overlapped on the planning CT for better visualization. (B) The model performance metric of the proposed EDOF model, three variants of EDOF model in the ablation study, and two comparison models.*

Figure 4(A) presents the normalized mean absolute scores (feature importance) of the *P* dosiomic and *Q* radiomic features derived from the trained EDOF model. As highlighted in the red box, three features exhibited significantly higher scores: (1) dosiomic *Intensity Mean*, which represents the average dose (directly convertible to Gy) within localized lung volumes (7×7×7 mm³), (2) radiomic *GLRLM-based Short Run Low Gray Level Emphasis* (*SRLGLE*), which emphasizes low-intensity regions (i.e., high air/tissue ratio) with significant structural heterogeneity, and (3) dosiomic *Intensity Mean Absolute Deviation* (*MAD*), which represents dose variation (convertible to Gy) within localized lung regions.

The upper panels of Figures 4(B-D) show the PDPs for these key features, illustrating the relationship between feature variations and model output. The solid line represents the mean values across the 5-fold validation, with shaded areas indicating standard deviations. The lower panels display histograms of feature distributions across the cohort of 72 patients, showing the frequency of specific feature values. Notably, regions in the PDPs with greater variability often correspond to histogram regions with lower feature counts.

1) Figure 4(B) displays the PDP for the dosiomic *Intensity Mean*. RP risk is minimal at lower doses (up to ~5 Gy). Beyond this threshold, the risk increases markedly, following a nearly linear trend from 8 Gy to 30 Gy. The risk gradually flattens in the 30-48Gy range.
2) Figure 4(C) displays the PDP for the radiomic *GLRLM-based SRLGLE*, showing a complex relationship with RP risk. At low SRLGLE values (0.000–0.006), RP risk is high with notable variability. Risk decreases at SRLGLE values between 0.006 and 0.010, then increases as SRLGLE rises from 0.010 to 0.025, after which RP risk declines steadily beyond 0.025. This pattern suggests a nuanced influence of structural heterogeneity on radiation response.
3) Figure 4(D) displays the PDP for the dosiomic *Intensity MAD*. RP risk increases sharply as variation approaches 1 Gy but plateaus beyond this point.

Figure 5 illustrates three representative cases (Case 1-3) accurately predicted by the EDOF model. The first row of each case presents the planning CT, planning dose distribution, ground truth RP volume, and the predicted RP volume from the EDOF model. Ground truth and predicted RP volumes are marked in red and overlaid on the planning CT. In Cases 1 and 2, both patients received doses exceeding 45 Gy in the upper left lung region, subsequently developing RP in the corresponding areas within three months post-treatment. The EDOF model accurately identified these high-risk RP regions, demonstrating strong visual agreement with the actual RP contours. In Case 3, despite receiving a high lung dose (>45 Gy), no RP was observed on follow-up CT scans. The EDOF model correspondingly predicted minimal high-risk regions, with only scattered high-risk pixels. The second row of these 3 cases shows the spatial distribution of three key dual-omic features, visualized using a jet colormap and overlapped on lung CT. The dosiomic *Intensity Mean* map highlights regions of high locoregional dose intensity, while the dosiomic *Intensity MAD* map identifies areas with significant dose falloff. The radiomic *GLRLM-based SRLGLE* feature map reveals high values in heterogeneous lung structure with high air/tissue ratio. For better visualization, the SRLGLE map was segmented into four value ranges, i.e., 0.000–0.006, 0.006–0.010, 0.010–0.025, and >0.025, as shown in the third row.

Figure 6 showed two cases (Case 4 and 5) where the EDOF model encountered prediction challenges. In Case 4, the actual RP volume extended beyond the high-dose planning region, significantly larger than model's prediction. In Case 5, RP occurred in a low-dose region far from areas of high radiation exposure. In both cases, the EDOF model failed to accurately predict the RP volumes.

Table I. the model performance metric of the proposed EDOF model, three variants of EDOF model in the ablation study, and two comparison models. The prediction performance was evaluated with five-fold cross validation (mean ± standard deviation), and "*" marks the statistically significant difference compared to our EDOF model.

| Model | AUC | Accuracy | Sensitivity | Specificity |
| --- | --- | --- | --- | --- |
| The proposed EDOF | **0.95±0.01** | 0.93±0.02 | **0.81±0.05** | 0.93±0.01 |
| EBM with radiomic filtering | 0.68±0.07* | 0.89±0.03* | 0.01±0.01* | 0.85±0.02* |
| EBM with dosiomic filtering | 0.94±0.01* | 0.90±0.03* | 0.71±0.03* | 0.92±0.02* |
| GBT with dual-omic filtering | 0.95±0.01 | 0.93±0.03 | 0.79±0.10* | 0.93±0.03 |
| U-Net++ with 3D CT and dose | **0.92±0.06*** | **0.98±0.00*** | 0.00±0.00* | **1.00±0.00** |
| DVI Model (patient-wise prediction) | 0.69±0.08 | 0.6±0.08 | 0.39±0.10 | 0.73±0.10 |

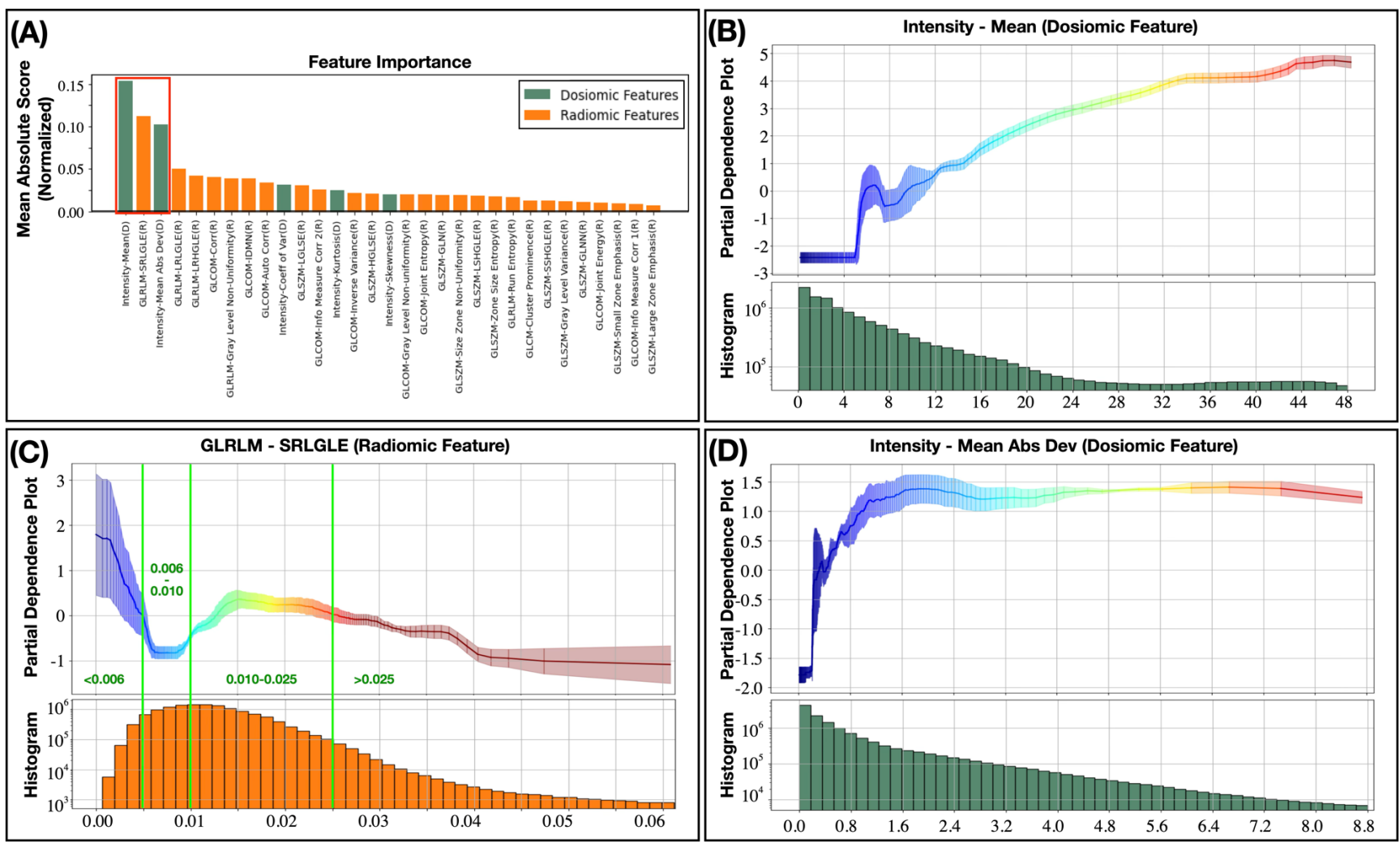

*Figure 4. (A) Normalized EBM mean absolute scores (feature importance) derived from the trained EDOF model, highlighting three key features (as in the red box) with significantly high scores among the feature sets. (B-D) PDPs and feature distribution histogram for the three key features.*

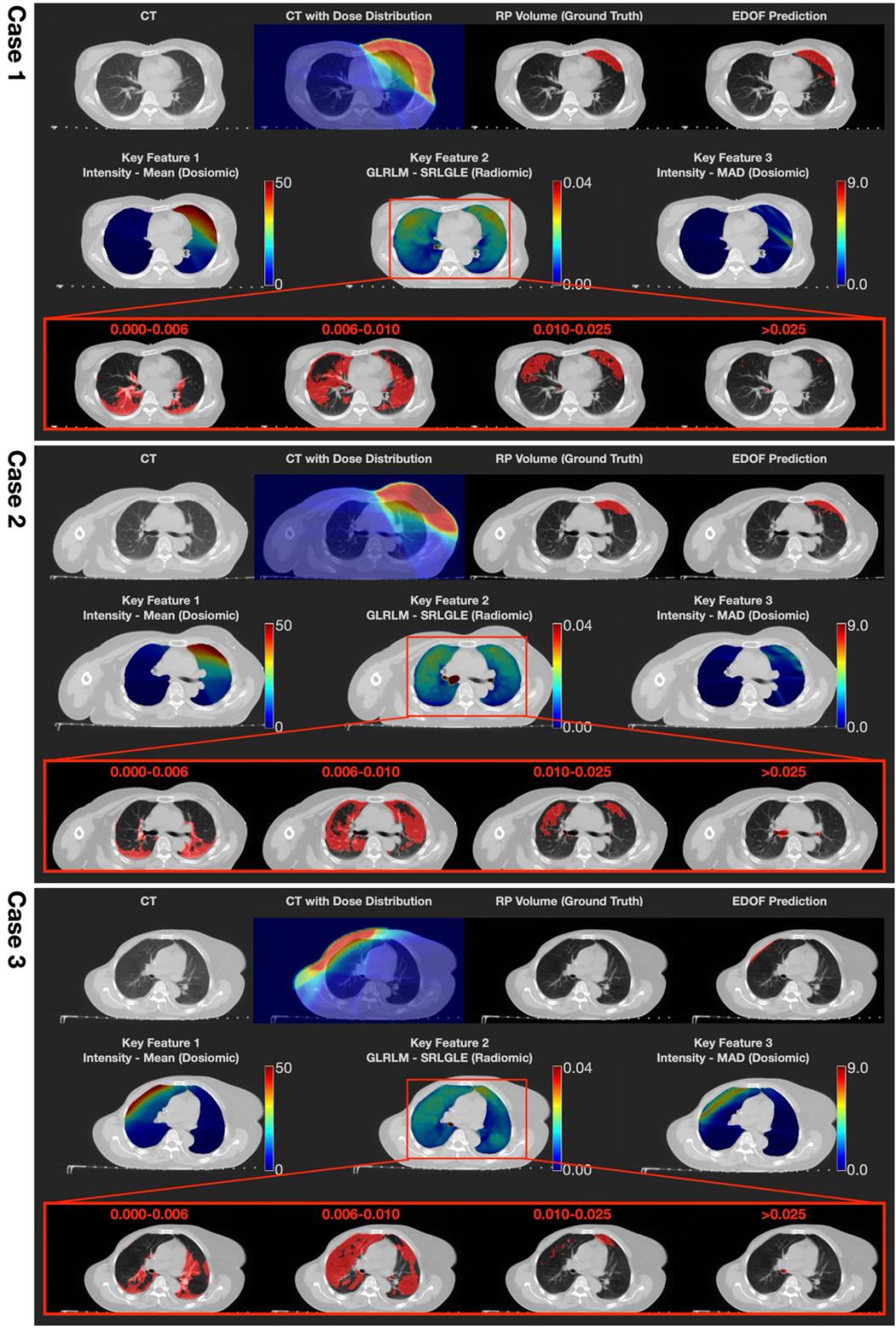

*Figure 5. Representative cases (Case 1-2) demonstrating the EDOF model's predictive performance. For each case, the first row shows the planning CT, dose distribution, identified RP volume, and predicted RP volumes; the second row shows the filtered maps of three identified key features; the third row shows four different feature ranges of the radiomic GLRLM-SRLGLE feature map.*

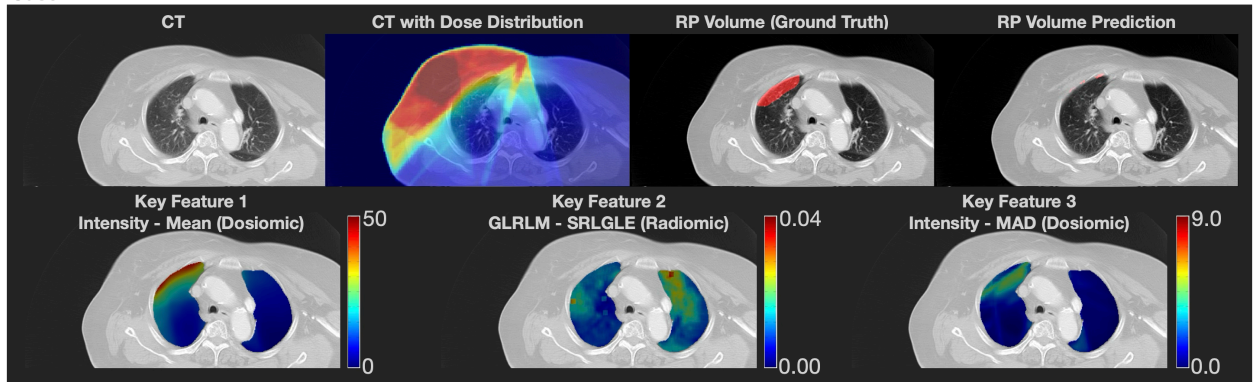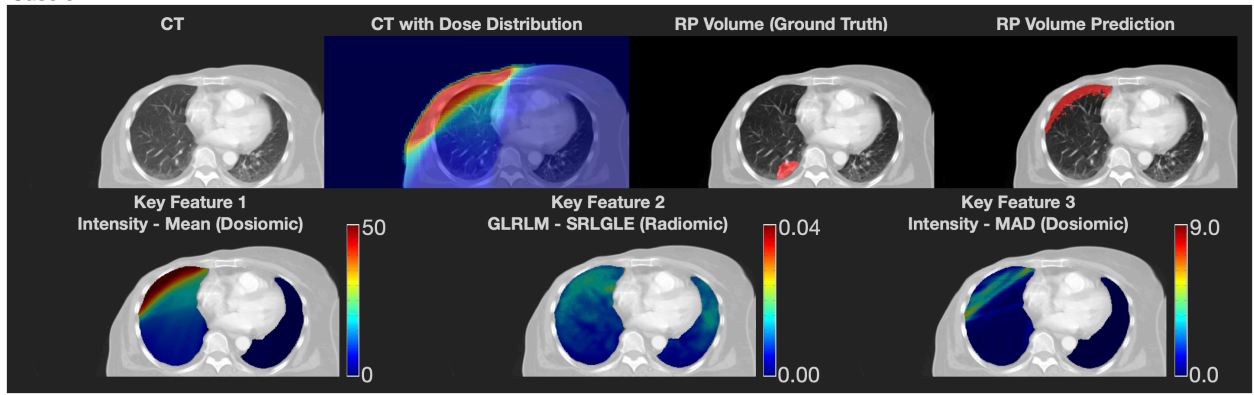

*Figure 6. The planning CT, planning dose distribution, identified RP volume (ground truth), and the predicted RP volume from the EDOF model for two patients (Case 3 and 4).*

## 5. Discussion

The management of radiation-induced pneumonitis in IMRT remains a significant challenge for breast cancer patients. The accurate prediction of locoregional RP risk before the treatment delivery is critical in supporting safer, personalized treatment planning. We hypothesis that pre-treatment locoregional CT radiomic textures, combined with locoregional dose distribution intensities from treatment planning, are intrinsically associated with post-treatment RP occurrence. To test this hypothesis, we developed a dual-omic filtering workflow with two key components: (1) radiomic filtering, which extracts locoregional texture features from segmented lung CT using a 3D sliding window, and (2) dosiomic filtering, which captures locoregional intensity features from planning dose distribution. Together, these features construct a spatially encoded feature vector for each voxel coordinate within the lung. The EBM was subsequently employed to predict RP risk on a voxel-wise basis. Collectively, the developed EODF demonstrated high predictive performance in a cohort of 72 breast cancer patients.

The EBM mean absolute score identified three key features contributing to RP risk: dosiomic *Intensity Mean*, radiomic *GLRLM-based SRLGLE,* and dosiomic *Intensity MAD.* The PDPs of these features provide detailed explanations of EBM decision-making process.

- The dosiomic *Intensity Mean* represents the localized mean dose. The PDP reveals a sharply increasing RP risk starting at the 5 Gy threshold. This directly supports the clinical practice of monitoring V5Gy as an early RP risk indicator. Beyond this threshold, PDP showed that RP risk increase almost linearly between 8-32 Gy. These findings validate the clinical monitoring of V10Gy, V20Gy and V30Gy, as RP risk rises rapidly within this range, while also offering a finer-grained understanding of dose-response relationships across varying dose levels. To our knowledge, this is the first study to provide such detailed, model-based explanations of dose intensity's impact on RP risk.
- The dosiomic *Intensity MAD* captures locoregional dose variability. The PDP showed that RP risk sharply increased up to a variability threshold of 1 Gy. Beyond this point, additional variability does not significantly elevate RP risk independently. Since irradiated lung volumes in breast cancer IMRT are typically limited, regions with *Intensity MAD*<1 Gy often correspond to lower-dose areas farther from the primary radiation target. In contrast, higher values are often associated with regions closer to the treatment field. This observation aligns with clinical findings that RP risk is predominantly influenced by cumulative high-dose exposure rather than variability within lower-dose regions.

- The radiomic GLRLM-based SRLGLE captures structural heterogeneity in lung tissue by evaluating the distribution of consecutive intensity values. This feature emphasizes tissue structural heterogeneity with high air/tissue ratio. The complex PDP pattern can be analyzed within the following four ranges:
  - Low SRLGLE Values (0.000–0.006): The model associates these values with elevated RP risk. As shown in Figure 5, these low SRLGLE values often presented in the regions near lung walls or peripheral lung edges. Due to the kernel-based radiomic filtering method used in this study, feature calculations in boundary regions often involve padding, which can lead to artificially low SRLGLE values. As the RP is often observed near lung edges, our model incorrectly associates the higher predicted risk in these regions. However, the increased variability in the PDP indicates uncertainty in reliably capturing RP risk for these areas.
  - Moderate SRLGLE Values (0.006–0.010): A local minimum in RP risk is observed in this range. This range typically represents lung regions with higher structural homogeneity and tissue density. While the EDOF model does not explicitly encode kernel locations, these moderate feature value often appear in the posterior lung, as shown in Figure 5. These effects may be attributed to gravitational forces during supine CT imaging, which increase tissue density and heterogeneity in posterior regions. These areas often farther from primary radiation targets in breast cancer IMRT, which exhibit reduced RP probability. Additionally, the mechanical stability and connective tissue in these regions enhance resilience to radiation-induced injury[59]. As shown in Figure 5, despite Case 3's upper left lung receiving high radiation exposure (>45 Gy), the EDOF model accurately predicted low RP risk.
  - High SRLGLE Values (0.010–0.025): RP risk increases as SRLGLE values rise in this range. These range typically correspond to lung regions characterized by heterogeneous, low-density structures such as irregular distributions of alveoli, bronchioles, and blood vessels. These regions are particularly vulnerable to radiation damage due to their delicate architecture, high vascularization, and limited

- regenerative capacity. The thin walls of alveoli, the primary sites of gas exchange, are especially susceptible to disruption by radiation, leading to inflammation and compromised lung function. Furthermore, radiation-induced damage to endothelial cells lining the lung's blood vessels can result in vascular leakage and exacerbate inflammation[60,61]. As illustrated in Cases 1 and 2, lung areas with SRLGLE values in this range overlap with high-dose regions, exhibiting elevated RP risk.
  - Ultra-High SRLGLE Values (>0.025): RP risk decreases and stabilizes beyond this threshold. As shown in Cases 1–3, very limited lung volume fall into these regions.

These findings align well with prior research on pre-treatment RP prediction[62]. The dosiomic feature *Intensity Mean* consistent with established clinical dose-volume metrics such as V5Gy, V20Gy, and V30Gy. Studies have shown that limiting lung exposure to low and moderate doses (V5Gy≤30% or V20Gy≤20%) significantly reduces RP incidence[63,64], while controlling high-dose constraints (NTDmean≤31.8 Gy, or MLD≤13.5Gy) minimizes severe RP rates[65]. Our findings on radiomic features also consistent previous studies employing CT-based metrics for RP characterization. Luna et al. demonstrated that increased entropy and reduced uniformity in irradiated lung regions are strong predictors of RP[66], while Krafft et al. reported improved RP prediction accuracy when radiomics were incorporated into models (AUC increased from 0.51 to 0.68)[67]. Additionally, vascular and textural features, such as increased contrast and reduced homogeneity, have been linked to inflammatory changes that drive RP development[26,68]. To our knowledge, this is the first study to provide a data-driven, spatially encoded perspective on RP risk at the voxel level, offering precise and explainable predictions of high-risk regions. The obtained explanation successfully proves that our model is driven by (1) features that are appropriate for clinical practice and (2) decisions that are clinically defensible. By integrating clinical guidelines with voxel-level predictions, the EDOF model is expected to support clinicians in making individualized adjustments to radiation delivery, ultimately enhancing patient safety and treatment outcomes.

The PDPs in our EDOF model highlight the complementary roles of dosiomic and radiomic signatures in assessing RP risk. Ablation studies also confirmed this observation. The radiomic filtering-only EBM exhibited poor performance as it lacks information on radiation exposure, a key determinant of RP risk. The dosiomic filtering-only EBM showed improved performance but remained inferior to our EDOF model, as it failed to incorporate the lung textural characteristics that influence regional RP risk. The GBT model with dual-omics filtering achieved slightly lower performance than the EDOF model. While GBT is a state-of-the-art ML classifier capable of capturing complex data patterns, it is more prone to overfitting, especially with noisy or limited data. The U-Net++ model required significantly larger model parameters and more complex nested layers. Due to the imbalanced label distribution, with RP volumes constituting a small fraction of the lung, U-Net++ showed severely compromised performance in capturing voxel-wise RP volumes (sensitivity=0.00±0.00). Furthermore, U-Net++ demanded substantially greater computational resources, with training times up to 20 times longer than the EDOF model. In the traditional "black boxes" models such as U-Net++ or GBT, explainability is challenging because the complex interactions and dependencies among features are distributed across numerous layers or trees. While post-hoc explainability methods, such as SHAP (Shapley Additive Explanations) values, can offer some insights into feature importance, they are often unable to directly capture the complex, feature-specific dependencies. A patient-level RP prediction model based on classic DVI demonstrated limited performance. However, this model's results are not directly comparable to the voxel-wise EDOF model.

It is important to note that our EDOF framework not only identifies key contributors to RP risk but also clarifies the model's limitations and applicable scenarios. Cases 4 and 5 in the highlight several limitations of our EDOF model in RP prediction. In Case 4, our model failed to predict the actual RP volume that extends beyond the planning high-dose region. This discrepancy may be associated with a number of factors such as positioning variations during treatment fractions. In Case 5, RP developed in a low-dose region, suggesting that factors such as prior chemotherapy, surgery, variables, and genomic markers may also play roles. The PDPs of the identified three key features effectively convey to clinicians the specific scenarios in which the EDOF model may fail. This level of explainability is critical for building trust and confidence, ensuring the safe and effective integration of ML models into clinical workflows. It should also be noted that this feasibility study was conducted with a limited dataset of 72 patients, focusing primarily on technical development. A more comprehensive evaluation of the model's robustness will require larger, multi-institutional datasets. Future improvements to the EDOF framework may include integrating Cone-Beam CT data and detailed treatment histories, which could help address the limitations identified in Cases 4 and 5 and further enhance the model's predictive accuracy and clinical utility.

## 6. Conclusion

This study developed a novel EDOF model to predict locoregional RP regions for breast cancer patients by incorporating both radiomic and dosiomic feature analysis. The model successfully identified key dual-omic features and provided detailed explanation of their impact to RP risk. These advancements have the potential to improve clinicians' understanding of RP prediction mechanisms and enable more personalized radiation therapy strategies to reduce RP incidence and severity.

## Conflict of Interest

None associated with this work